\documentstyle[12pt]{article}

\newcommand {\be}{\begin{equation}}
\newcommand {\ee}{\end{equation}}
\newcommand{\bea}{\begin{eqnarray}}
\newcommand{\eea}{\end{eqnarray}}
\def\Sp{{\rm Sp\,}}

\begin{document}

\title {STATISTICAL ASPECTS OF NEVEU\\ AND SCHWARZ DUAL MODEL}

\author {M.I.~GORENSTEIN, V.P.~SHELEST,\\ Yu.A.~SITENKO, G.M.~ZINOVJEV}
\date{Received: March 16, 1973}
\maketitle

\begin{abstract}
\noindent
Statistical method is applied to Neveu and Schwartz model
for obtaining the average characteristics of the heavy resonances:
spin distribution, decay widths etc. The properties of the dual
model spectrum of states constructed of both commuting and
anticommuting operators are considered. In this case the spin
characteristics coincide with the results of the statistical
bootstrap model.
\end{abstract}

\newpage
\section
{Introduction}

        In this paper we investigate the average characteristics
of the Neveu and Schwarz dual model from the statistical view-point
[1---4]. The exponential rise of the density of hadron states at high
energies in the dual resonance model (DRM) creates the preconditions
for the efficiency of the statistical approach. Besides, there may
exist an essential connection [3] between the properties of the
resonance spectrum in the DRM and in the statistical bootstrap model
[5, 6].

        For some observable $A$ the average in the resonances with
fixed mass $M$ (i.e. microcanonical average) $\alpha(M^2)\equiv
\alpha(0)+\alpha'M^2=N$ is defined in dual resonance model to be [3]
\be
\langle A \rangle_N = {{\sum\limits_{R\in N}\langle R \vert A \vert R
\rangle}\over{\sum\limits_{R\in N}\langle R\vert R\rangle}} = {{{1\over{2\pi
i}}
\oint {dz z^{-N-1} {\rm Sp\,}(z^HA)}}\over{{1\over{2\pi i}}\oint {dz
z^{-N-1}
{\rm Sp\,} z^H}}},
\ee
where $H$ is the Hamiltonian operator in DRM, and the integration
contour in the complex $z$-plane envelopes the point $z=0$.

        In the paper [4] a new type of averaging has been proposed

\bea
{\langle\!\langle A \rangle\!\rangle_N}&=& {\sum\limits_{R\in
N}{|\langle 0\vert V(p)\vert R\rangle \vert^2\langle R\vert
A\vert R\rangle}\over\sum\limits_{R\in N}{\vert\langle 0 \vert V(p) \vert
R\rangle\vert^2}}=\nonumber\\
&=&{{1\over(2\pi i)^2}\oint {dx}\oint dy
x^{-N-1}y^{-N-1}\langle0\vert V(p)x^HAy^HV^+(p)\vert0\rangle\over
{1\over2\pi i} \oint dx\,x^{-N-1}\langle 0 \vert V(p)x^HV^+(p)\vert
0 \rangle},
\eea
where $V(p)$ is the vertex operator
in DRM. Every resonance state is contained in the average (2) with the
weight proportional to the square of the coupling constant with
the basic states of the model (conditionally $\pi$-mesons) $g_{\pi\pi
R}= \langle0\vert V(p)\vert R\rangle$. This provides an effective
accounting of the dynamical features of the reactions and allows
us to call (2) the dynamical average.

        In the papers [3, 4] the averages such as (1) and (2) were
considered in the framework of the Veneziano dual model. It is clear
that the statistical approach can be applied to the other possible
formulations of the dual model. We shall consider here the dual model
of Neveu and Schwarz which possesses some important theoretical
advantages comparing with the usual Veneziano model. In this model
the resonance states are generated by the infinite sequence of
operators of the usual Veneziano model

\be
\big[a^\mu_n,a^{\nu +}_{n'}\big] = -g^{\mu\nu}\delta_{nn'}
\ee
$$(n,n' = 1,2, ...)$$
and the new  infinite set of anticommuting operators

\be
\big\{b^\mu_m,b^{\nu +}_{m'}\big\} = -g^{\mu\nu}\delta_{mm'}
\ee
$$(m,m' = 1/2, 3/2, ...)$$
satisfying conditions $\bigl[{a^\mu_n,b^{nu +}_m}\bigr] = 0$. The role
of Hamiltonian is played by the operator

\be
H = H_a + H_b = -\sum^\infty_{n=1}na^+_na_n - \sum^\infty_{m=1/2}
mb^+_mb_m
\ee
and the vertex operator is defined as

\be
V(p)=V_a(p)V_b(p)=\exp\Big[-\sqrt2p\sum_n{{a^+_n}\over{\sqrt n}}
\Big]\exp\Big[\sqrt 2p\sum_n{{a_n}\over{\sqrt n}}\Big]p\sum_m(b^+_m+b_m).
\ee
The eigenstates of the operator $H$ are divided into two groups
with different quantities of $G$-parity [7]
$\bigl(G=(-1){\vphantom{\frac{1}{2}}}^{\sum_m
{b^+_mb_m}}\bigr)$:

\begin{enumerate}
\item[1)]
the states with eigenvalues $N={1\over2}, {3\over2}, ...$
composed of the odd number of ''$b$''-operators and an arbitrary number
of ''$a$''-operators, $G=-1$;
\item[2)] the states with eigenvalues $N=1,2, ...$ composed of the
even number of ''$b$''-operators and an arbitrary number of
''$a$''-operators,
$G=1$ (the state with $N=0$ is spurious [7]).
\end{enumerate}

        In the particle interactions the value of $G$-parity is conserved.
The $\pi$-meson is associated with the lowest physical state of the model
with the mass $m^2_\pi = -{1\over2}$.

        The low energy part of the spectrum in the Neveu-Schwarz model
is similar to the experimental data.

\section
{Statistical Averages}

        We shall start with considering the degeneracy of the resonance
level $N$, i.e. the number $d(N)$ of the solutions of

\be
{H|R_i\rangle}=N|R_i\rangle, (i=1,2,\ldots d(N)).
\ee
It is obvious that the degeneracy $d(N)$ can be presented in the
following form

\bea
&&\hspace{-1cm}d(N)={1\over2\pi i} \oint dz\,z^{-2N-1} {\rm
Sp\,}\big(z^{2H}\big)= {1\over2\pi i} \oint dz
z^{-2N-1}\times\nonumber\\
&&\hspace{-1cm}\times\prod^\infty_{n=1}(1-z^{2n})^{-4}
(1+z^{2n-1})^4 ={1\over2\pi i} \oint dz\,
z^{-2N-1}\theta^{-2}_2(0\vert\tau)\theta^{-2}_4(0\vert\tau)4z^{1/2},
\eea
where $\tau = -i {\ln z\over\pi}$, and $\theta_2$, $\theta_4$ are
the Jacobi $\theta$-functions [8]. The behaviour of (8) as $N$ goes
to infinity is determined by the quantity of the integrand at $z\simeq
1$, which is given by means of the imaginary Jacobi transformation
[8].  The integral (8) is calculated by employing the saddle-point
technique, and

\be
T={1\over\pi} \sqrt{N}
\ee
determines the saddle point $Z_0=e^{-{1\over{2T}}}$. We get

\be
d(N) \simeq {1\over2^4}N^{-7/4}\exp[2\pi\sqrt{N}].
\ee
Taking into consideration that $N \simeq \alpha'M^2$, we get the
expression for the hadron state density $\rho(M)$ at
$M\to\infty$

\be
\rho(M) \simeq {\rm const}\cdot
M^{-5/2}\exp\biggl[{M\over T_0}\biggr], \ee
where
$T_0={1\over2\pi\sqrt{\alpha'}}$ plays the role of Hagedorn limiting
temperature [5]. For $\alpha'\simeq1$~GeV$^{-2}$ we get $T_0\simeq
160$~MeV, and this result remarkably coincides with the prediction of
the thermodynamic model [5] based on the comparison with the
experiment.

        Let us calculate now the average number of the oscillator
excitations. According to (1), we have

\bea
&&\langle-a^+_na_n\rangle_N = {1\over d(N)2\pi i}\oint dz\,
z^{-2N-1} \Sp\big(-a^+_na_nz^{2H}\big) =\nonumber \\
&&={1\over d(N)2\pi
i}\oint dz\, z^{-2N-1}
{4z^{2n}\over1-z^{2n}}\prod^\infty_{k=1}\big(1-z^{2k}\big)^{-4}\big(1+z^{2k-
1}\big)^4.
\eea
At fixed $n$ and $N\to\infty$ we get

\be
\langle-a^+_na_n\rangle_N \simeq {4\over e^{n/T}-1},
\ee
where $T$ is given by (9).

        Analogous calculation for the number of ''$b$''-excitations
gives

\be
\langle-b^+_mb_m\rangle_N \simeq {4\over e^{m/T}+1}.
\ee
Expressions (13) and (14) are the equilibrium distributions of
Bose-Einstein and Fermi-Dirac, respectively. Thus, we have come
to the canonical distribution with the temperature $T$ given by (9).
The use of the saddle point technique while calculating the
integrals such as (1) enables us to fulfil the normal in
statistical mechanics transition from the microcanonical to the
canonical (Gibbs) ensemble. The canonical averages of the physical
values are defined as

\be
\langle A\rangle_T = {\Sp\big(e^{-H/T}A\big)\over \Sp e^{-H/T}}.
\ee
Let us continue considering the average characteristics of the
resonance spectrum using (15). The total mean number of the
oscillator excitations of ``$a$" and ``$b$'' type are equal,
respectively, to

\be
\langle-\sum_{n=1, 2,\ldots} a^+_na_n\rangle_T = \sum_n
{4\over e^{n/T}-1} \simeq 4T \ln T,
\ee
\be
\langle-\sum_{m=1/2, 3/2, \ldots} b^+_mb_m\rangle_T = \sum_m
{4\over e^{m/T}+1 } \simeq 4T \ln2 .
\ee
The canonical average of energy has to coincide with the energy
$N$ of the microcanonical ensemble and is given by the expression:

\be
\langle H \rangle_T=\langle H_a \rangle_T + \langle H_b \rangle_T
\simeq\sum_n{4n\over e^{n/T}-1} + \sum_m{4m\over e^{m/T}+1} =
{2\over3}\pi^2T^2 + {1\over3}\pi^2T^2.
\ee
In (16), (17), (18) we substitute the summation for the integration
which is reasonable for large $T$.

        Let us proceed to the question of spin distribution in the
Neveu and Schwarz dual model. In the case of Veneziano model this task
has been solved in the papers [2, 9]. The operator of the $z$-component
of spin is [10]
\be
L_z=L^a_z+L^b_z=i\sum^\infty_{n=1}(a^{x+}_na^y_n-a^{y+}_na^x_n) + i
\sum^\infty_{m=1/2}(b^{x+}_mb^y_m-b^{y+}_mb^x_m).
\ee
Choosing the basis of states to be diagonal in $L_z$ [2], we find
the characteristic function of the operator $L_z$ according to (15)

\bea
&&\hspace{-1cm}\langle e^{i\varphi L_z}\rangle_T =
\prod^\infty_{n=1}\Bigl[{
(1-e^{-n/T})^2\over 1- 2 \cos \varphi
e^{-n/T}+e^{-2n/T}}\Bigr]
\Bigl[ {1+2\cos\varphi e^{-{n-1/2\over T}}+e^{-{2n-1\over T}}
\over(1+e^{-{n-1/2\over T}})^2}\Bigr]=\nonumber\\
&&\hspace{-1cm}=
\theta^{-1}_1(v|\tau)\theta_2(0|\tau)\theta_3(v|\tau)\theta_4(0|\tau)
\sin\pi v,
\eea
where $v={\varphi\over 2\pi}$, $\tau={i\over 2\pi T}$ and $\theta_1,
\theta_2, \theta_3, \theta_4$ are the Jacobi $\theta$-functions. For
$T\to\infty$ we apply the imaginary Jacobi transformation
for the evaluation of (20), and get

\be
\big\langle e^{i \varphi L_z}\big\rangle_T \simeq {2\pi T \sin{\varphi
\over2}\over\sinh \pi \varphi T}.
\ee
By differentiating (21) we find the average value

\be
\big\langle L_z^2\big\rangle_T = - {d^2\over d\varphi^2}
\big\langle e^{i \varphi L_z}\big\rangle_T \big|_{\varphi = 0} \simeq
{\pi^2T^2\over3}={N\over3}.
\ee

The distribution in $l_z$ is the following

\bea
\sigma_N(l_z)={1\over 2\pi}\int^\pi_{-\pi} d\varphi e^{-i \varphi l_z}
\big\langle e^{i \varphi L_z}\big\rangle_T \simeq
{1\over 4T} {1\over \cosh^2\big({l_z\over 2T}\big)}.
\eea
The integral in (23) is calculated by contour integration enclosing
the contour in the lower half-plane and explicitly performing the
sum over the pole residues. Using (23) one can easily obtain the
spin distribution of the resonances

\be
\rho_N(l) \simeq -(2l+1){d\over dl_z}\sigma_N(l_z)\big|_{l_z=l} = {1\over
4T^2}
{(2l+1)\sinh\big({l\over2T}\big)\over \cosh^3\big({l\over2T}\big)},
\ee
which can give further information about the spin structure of the
resonance spectrum. Let us write out the spin characteristics of the
resonances constructed of ''$b$''-operator only. The characteristic
function of the operator $L^b_z$ is

\be
\big\langle e^{i \varphi L_z^b}\big\rangle_T=\prod^\infty_{n=1}
{1+2\cos\varphi e^{-{(n-1/2)\over T}}+e^{-{(2n-1)\over T}}\over
(1+e^{-{(n-1/2)\over T}})^2}={\theta_3(v|\tau)\over \theta_3(0|\tau)}
\mathop{\simeq}_{T\to\infty} e^{-{\varphi^2\over2} T}.
\ee

        From (25) we get

\be
\big\langle(L_z^b)^2\big\rangle_T= -{d^2\over d\varphi^2}
\big\langle e^{i \varphi L_z^b}\big\rangle_T\big|_{\varphi=0}\simeq
T={\sqrt N\over \pi}.
\ee
We see that for $N\to\infty$ the ``$b$''-oscillator contribution
into spin becomes asymptotically negligible, whereas their contribution
into the energy makes up $N\over3$ according to (18). The distribution
in $l_z$ for the ``$b$'' oscillators has the Gaussian form

\be
\sigma_N^b(l_z)={1\over2\pi}\int^\pi_{-\pi} d\varphi e^{-i \varphi l_z}
e^{-{\varphi^2\over 2}T}\simeq
{1\over\sqrt{2\pi T}} e^{-{l_z^2\over 2T}}.
\ee

\section
{Dynamical Averages}

        The dynamical averaging (2) in Neveu-Schwarz model is
performed over the states with integer values of $N$ (it is necessary
for the $G$-parity conservation). The characteristic function of the
operator $L_z$ is given by

\be
\langle\!\langle e^{i\varphi L_z}\rangle\!\rangle_N = {{1\over2\pi i}\oint
dx
x^{-N-1}
\langle\pi|V(p)x^He^{i\varphi L_z}V^+(p)|\pi\rangle\over {1\over 2\pi i}
\oint dx x^{-N-1}\langle\pi|V(p)x^HV^+(p)|\pi\rangle},
\ee
where $|\pi\rangle = kb^+_{1/2} | 0 \rangle$ is pion state [7].
Using kinematical relations in the c.m.s. $(\vec p + \vec k =
0)$ and the mass-shell conditions $(p^2 = k^2 = m^2_\pi =
- {1\over 2})$, we obtain

\bea
&&\langle \pi|V(p)x^He^{i\varphi
L_z}V^+(p)|\pi\rangle=\big[(1-x)^{-1+2(p_x^2+
p_y^2)(1-\cos\varphi)}\big]\times\nonumber\\
&&\times\big[{1\over4} {x\over 1-x}+{N^2\over4}(1-x)+(p_x^2+p_y^2)(Nx-
{x\over 1-x})(1-\cos\varphi)\big].
\eea

        From (28) and (29) we find

\be
\langle\!\langle L_z^2 \rangle\!\rangle_N=-{d^2\over
d\varphi^2}\langle\!\langle
e^{i\varphi L_z}\rangle\!\rangle_N\big|_{\varphi=0}\simeq 2(p_x^2+p_y^2)
[\ln (N-1)+C+0({1\over N})],
\ee
where $C = 0,577 \ldots$ is the Euler constant.

               Choosing the momenta in the c.m.s. along the $x$-axis, for
(30)
we get

\be
\langle\!\langle L_z^2 \rangle\!\rangle_N \simeq {N\over2}\ln N.
\ee
Also, it is easy to find the contribution of ``$b$''-oscillators

\be
\langle\!\langle (L_z^b)^2 \rangle\!\rangle_N \simeq 4(p_x^2+p_y^2) \simeq
N.
\ee
The average numbers of the excitations are

\be
\langle\!\langle-a^+_na_n\rangle\!\rangle_N={1\over n}\big(1-{n\over
N}\big);
{\rm for\,}  n<N,
\ee
\be
\langle\!\langle-b^+_mb_m\rangle\!\rangle_N={1\over N}; {\rm for\,}
{1\over2}<m<N.
\ee

\section
{Resonance Decay Widths}

         Let us take up now a heavy resonance decay within the framework
of the Neveu and Schwarz model. According to [3], the average decay
width of the resonance from $N_1$-level into the resonance from $N_2$-
level and $\pi$-meson is given by

\be
\langle\Gamma\rangle={1\over d(N_1)(2\pi i)^2}\oint dx\oint dy\,
x^{-2N_1-1}y^{-2N_2-1}\Sp\big(x^{2H}V(p)y^{2H}V^+(p)\big).
\ee
The trace in (35) has the following form

\bea
&&\hspace{-2cm}\Sp\Big(x^{2H}V(p)y^{2H}V^+(p)\Big) =
\Big[(1-x^2)^{-1}\prod^\infty_{n=1}
(1-(xy)^{2n})^{-4}\times\nonumber\\
&&\hspace{-2cm}\times {(1-(xy)^{2n})^2\over(1-x^2(xy)^{2n})(1-{1\over
x^2}(xy)^{2n})}\Big]
\Big[{1\over 2}\prod^\infty_{n=1}
(1+(xy)^{2n-1})^4\sum^\infty_{k=1}{x^{2k-1}+y^{2k-1}\over
1+(xy)^{2k-1}}\Big].
\eea
To calculate the integral (35), we introduce the variables $\omega = xy$
and $x$. The main contribution to the integral (35) with large $N_2$,
$N_1 - N_2$ is given by the region $\omega \simeq 1, x \simeq 1$.
Expressing the integrand in terms of the Jacobi $\theta$-functions, we
obtain for $\omega\simeq 1$

\be
\prod^\infty_{n=1}{(1-\omega^{2n})^2\over(1-x^2\omega^{2n})
(1-{1\over x^2}\omega^{2n})} \simeq {\pi \sinh(\ln x)\over\ln \omega
\sin ({\pi \ln x\over \ln \omega})}\exp\big[{\ln^2x\over \ln \omega}\big],
\ee
\be
\sum^\infty_{k=1}{x^{2k-1}+\big({\omega\over x}\big)^{2k-1}\over
1+\omega^{2k-1}} \simeq - {\pi\over 2\ln \omega \sin\big({\pi \ln x
\over \ln \omega}\big)}.
\ee
The integration over $\omega$ is performed by employing the saddle-point
technique with the result

\bea
&&\langle\Gamma\rangle = {d(N_2)\over d(N_1)} {1\over 2\pi i}
\oint dx\,x^{-2N_1+2N_2-1}(1-x^2)^{-1} f(x,\omega_0);\nonumber\\
&&f(x,\omega_0) \mathop\simeq_{x\simeq 1}\big({\pi\over 2\ln
\omega_0}\big)^2
{(-\sinh \ln x)\over \sin^2 {\pi \ln x\over \ln \omega_0}},
\eea
where $\omega_0 =\exp\left[-{\pi\over 2\sqrt{N_2}}\right]$ is the saddle
point. The remaining integral is easily calculated at $N_1-N_2 >>
\sqrt {N_2}$ which corresponds to the large energy of the emitted
$\pi$-meson. In this case ${\ln x\over \ln {\omega_0}} \simeq
{\sqrt {N_2}\over N_1-N_2} \simeq 0$, and so we get

\be
\langle\Gamma\rangle = {d(N_2)\over d(N_1)}{N_1-N_2\over 2}.
\ee
Expression (40) may be written in terms of the decaying resonance
mass $M_1$ and the $\pi$-meson energy $E$

\be
\langle\Gamma\rangle \simeq \big(1-{2E\over M_1}\big)^{-7/4}EM_1
\exp\big[- {E\over{T_0\over2}(1+\sqrt {1-{2E\over M_1}})}
\big].
\ee
We see that the meson energy spectrum has the Boltzman distribution
form with the effective temperature $T_{eff} = {T_0\over 2} \bigl(1+
\sqrt {1-{2E\over M_1}}\bigr)$. Such a dependence coincides exactly
with the results of the statistical bootstrap model [12] (note that
we do not take into account the whole chain of the successive decays,
the consideration of which changes the value of $T_{eff}$ [12]).

        A considerable difference of (41) from the analogous result
obtained in the Veneziano model with $\alpha(0) \not=1$ [3] consists
in the linear growth of the decay width with $M_1$ increasing (such a
behaviour would take place also in the Veneziano model with
$\alpha(0) = 1$).

\section
{Discussion}

        In conclusion we would like to discuss some interesting, as
it seems to us, aspects of our consideration:

        1. The average resonance spectrum characteristics in the
Neveu-Schwarz model for $N \rightarrow \infty$ are similar, as a whole,
to those obtained in the Veneziano model [2, 3, 4]. This allows us
to hope that statistical approach grasps the main features of the dual
dynamics independent of the choice of the specific model.

        2. The average decay width has the same dependence on the
$\pi$-meson energy as in the statistical bootstrap model and grows
linearly as $M\rightarrow\infty$. Concerning the total width dependence
on the hadron mass, the comparison is meaningless, since no decay widths
but the relative probabilities are considered in the statistical
bootstrap and the total probability is normalized to be equal unit.

        3. In the Neveu-Schwarz model there exists an infinite
sequence of the gauge relations which allows one to remove the unphysical
states. In order to obtain the number of physical states, we must add
the factor $(1-z)\prod^\infty_{n=1}(1-z^{2n})(1+z^{2n-1})^-1$ to the
integrand (8) (the same factor as for calculating the planar loop)
[13, 11]. The number of physical states is given by [13]

\be
d^D_{Phys}(N) = d^{D-1}(N)-d^{D-1}(N-{1\over2}),
\ee
where $D$ is the dimension of the oscillator operators (in our case
$D = 4$). For large $M$ the density $\rho^{(M)}_{Phys}$ equals

\be
\rho_{Phys}(M) \simeq {\rm const} M^{-3} \exp[\pi\sqrt{3\alpha'}M].
\ee
The removal of unphysical states can be easily taken into account
in our calculations. For example, one must substitute (42) for $d(N_1)$
and $d(N_2)$ in the expression (40), which results in the alteration of
the quantity of $T_0$ in (41). So, the partial widths of the decay with
$\pi$-meson emission, which have the form $\langle\Gamma\rangle \sim
{d(N_2)\over d(N_1)}$, do not essentially change as the removal of
unphysical
states is performed. However, the situation may be quite different while
cosidering the general case of decay $R_1\to R_2+R_3$ where one expects
the results such as  $\langle\Gamma\rangle \sim {d(N_2)d(N_3)\over
d(N_1)}$.
Here the decay characteristic will be crucially dependent of the value of
power of M in the expressions (11), (43). Note that the case of the power
value equal to $-3$ (regarding the physical states only) coincides with
the results of the investigations [14, 15] and makes the decay into one
heavy and one light ($\pi$-meson) particles dominant, while the case of
the value equal to $-5/2$ (all states are taken into account) corresponds
to
the primary formulation of Hagedorn [5] which predicts the decay into
particles of approximately equal masses.

        4. Our calculations allow us to obtain the average characteristics
of the resonance spectrum constructed of ''$b$'' operators only.

        The question of building the dual theory with the same hadron state
density as in the statistical bootstrap has been considered in the recent
paper [16]. The author concludes that the construction of resonance states
with the only Fermi operators represents one of the most preferable
possibilities. In this case using calculations (26) and (27) we find that
also the average spin and the spin distribution coincide with that of the
statistical bootstrap theory with spin [17].

We thank B. V. Struminsky and V. A. Miransky for useful discussions.

\end{document}